%% file: main.tex
\documentclass{ifacconf}

\usepackage{ifpdf}
\usepackage{graphicx} 
\usepackage{float}  % Para controle de objetos flutuantes
\usepackage{caption}  % Para legendas de figuras e tabelas
\usepackage{subcaption}  % Para sublegendas
\usepackage{mathptmx,amsmath}  % Usa a fonte Times New Roman
\usepackage{amssymb}
\usepackage{listings}
\usepackage{tikz}                 % Pacotes para desenhar usando latex
\usepackage{cuted}                % Pacote para criar ambientes de figura sobre duas colunas
\usepackage{booktabs}  % Para tabelas de alta qualidade
\usepackage{algorithm, balance}
\usepackage[ruled,vlined,algo2e]{algorithm2e}

\usepackage{cite}
\usepackage{algorithmic}
\usepackage{fixltx2e}
\usepackage{stfloats}

\usepackage{amssymb}

\usepackage{etoolbox} % Para \AtBeginEnvironment

\usepackage{url}
\usepackage{etoolbox}

\def\GR#1{{\color{black}#1}}

\def\BibTeX{{\rm B\kern-.05em{\sc i\kern-.025em b}\kern-.08em
    T\kern-.1667em\lower.7ex\hbox{E}\kern-.125emX}}

\newtheorem{defi}{\textbf{Definition}}
\AtBeginEnvironment{defi}{\setlength{\parindent}{0pt}}

\newtheorem{myexmp}{\bf Example}
\AtBeginEnvironment{myexmp}{\setlength{\parindent}{0pt}}

\usepackage{graphicx}      % include this line if your document contains figures
\usepackage{natbib}        % required for bibliography
\begin{document}
\begin{frontmatter}

\title{Model Predictive Supervisory Control for Hierarchical and Distributed UAS Traffic Management
\thanksref{footnoteinfo}}

\thanks[footnoteinfo]{This study was supported in part by the Brazilian agencies CAPES through the Academic Excellence Program (PROEX) – Finance Code 001, CNPq under grants 317058/2023-1 and 422143/2023-5, FAPEMIG under fellowship APQ-06580-24, and in part by Petrobras/ANP under Grants 2023/00494-5 and 2023/00643-0.}

\author[First]{Matheus P. Loures} 
\author[First,Second]{Guilherme V. Raffo}
\author[First,Second]{Patrícia N. Pena}

\address[First]{Graduate Program in Electrical Engineering, Universidade Federal de Minas Gerais, Belo Horizonte, MG, Brazil {\tt \{mploures,raffo,ppena\}@ufmg.br}}
\address[Second]{Department of Electronics Engineering, Universidade Federal de Minas Gerais, Belo Horizonte, MG, Brazil}

\begin{abstract}                
This work proposes a hierarchical Model Predictive Supervisory Control (MPSC) framework for multi-agent systems with shared resources. MPSC integrates receding-horizon cost-optimal control with Supervisory control theory (SCT) based supervision that enforces safety, nonblockingness, and resource exclusivity. Scalability arises from hierarchical and scalable supervisor and automaton templates, enabling distributed execution without monolithic synthesis. Using this framework, this work develops an urban Unmanned aircraft system Traffic Management (UTM) model. The model supports pickup-and-delivery missions under time-varying demand efficiently.
\end{abstract}


\end{frontmatter}

\section{Introduction}
\input{sections_V2/01Introducao}
\section{Preliminaries}
\input{sections_V2/02Preliminaries}
\section{Problem Statement}
\input{sections_V2/03Problema}
\section{Model Predictive Supervisory Control} 
\input{sections_V2/04Metodologia}

\section{Case Study: Structured Urban Airspace}
\input{sections_V2/06Estudo}
\section{Conclusion}
\input{sections_V2/07Conclusao}

%\balance
\bibliography{ifacconf}

\end{document}

%% file: sections_V2/01Introducao.tex
The popularization of unmanned aerial vehicles (UAVs) has expanded beyond military and academic niches into everyday services, such as logistics, critical infrastructure inspection, environmental monitoring, and urban public services. The shared low-altitude airspace, however, constitutes a physical bottleneck. Consequently, Unmanned Aircraft Systems Traffic Management (UTM) requires conflict-free routing, time-constrained scheduling, ad hoc communication, distributed decision-making, and autonomous fault response.

Several global UTM initiatives emphasize dynamic authorization, robust communication, and real-time fleet monitoring. Examples include NASA's U.S. trials \citep{aweiss2018unmanned}, Europe's U-space framework for very-low-level (VLL) operations \citep{undertaking2017european}, China's low-altitude route networks \citep{xu2020recent}, and India's Digital Sky \citep{yadav2021uav}. In Brazil, VLL airspace (60–150 m) is under full state sovereignty, with geofences representing no-fly zones and safe routes \citep{ICA_100_40_Brasil}. However, advanced UTM development remains crucial for safely  carrying out complex operations within this space.

 Dense urban UTM operations require formal coordination methods capable of representing multi-agent interactions, sha\-red resources, and operational constraints. Supervisory control theory (SCT) provides a formal basis for synthesizing minimally restrictive supervisors that regulate discrete-event systems (DES). It achieves this by disabling a subset of events to satisfy formal specifications, thereby  ensuring  critical  properties like nonblocking behavior and controllability \citep{ramadge1987supervisory}. Prior SCT-based studies have addressed multi-agent coordination through distributed motion control \citep{roszkowska2013distributed}, scalable swarm robotics \citep{lopes2016supervisory}, and hierarchical safe navigation \citep{dulce2022distributed,vilela2022hierarchical}. Recent work has also combined SCT with model-predictive and multi-objective scheduling mechanisms for real-time UAV coordination \citep{loures2025adaptive}.

Building on this foundation, this paper proposes a hierarchical UTM framework structured around a model predictive supervisory control (MPSC) process, in which UTM-level supervisory restrictions define admissible airspace behaviors and UAV-level predictive optimization selects feasible local actions. This work integrates DES-based supervisory restrictions, receding-horizon decision-making, and MILP-based MPSC within a hierarchical UTM architecture for multi-UAV coordination, so that optimization is performed only over behaviors admitted by the supervisory control layer.

\vspace{-0.2cm}

%% file: sections_V2/02Preliminaries.tex
\label{sec:preliminaries}

This section introduces the fundamental concepts of DES and SCT that underpin our methodology, along with the automata-based prediction model representation used in the MPSC.
\vspace{-3mm}
\subsection{Discrete Event Systems and Supervisory Control}
\vspace{-3mm}
Discrete Event Systems (DES) are dynamic systems where state changes through instantaneous event-driven transitions, suitable for sequencing and synchronization~\citep{cassandras2008introduction}. An alphabet $\Sigma$ is a finite set of symbols. A string is a finite sequence of events, and the set of all possible sequences forms the Kleene closure $\Sigma^*$. A language is any subset of $\Sigma^*$. Deterministic finite automata (DFA) represent DES as $G = (Q, \Sigma, \delta, q_0, Q_m)$: $Q$ is the set of states, $\delta: Q \times \Sigma \rightarrow Q$ is the transition function, $q_0$ is the initial state, and $Q_m$ is the set of marked states. The extended transition function $\delta : Q \times \Sigma^{\ast} \rightarrow Q$ satisfies $\delta(q,\sigma s)=q'$ given $\delta(q,\sigma )=q_1$ and $\delta(q_1,s)=q'$. Generated language $\mathcal{L}(G)=\lbrace s\in \Sigma^{\ast}\mid\delta(q_0,s)\in Q \rbrace$ and marked language $\mathcal{L}_m(G)=\lbrace s \in \Sigma^{\ast}\mid\delta(q,s)\in Q_{m} \rbrace$ describe the system's behavior.The parallel composition of two automata $G_1||G_2$ produces a third DFA that synchronizes the behaviors of $G_1$ and $G_2$ in the common events. To incorporate performance criteria, a DFA is extended to a multi-weighted automaton based on \citep{fahrenberg2011energy}.
\begin{defi}\label{def:weighted}\citep{loures2025adaptive}
A multi-weighted automaton $G = (Q, \Sigma, \delta, q_0, Q_m, W)$ is a DFA with $W: Q \to \mathbb{R}^k$ that assigns a $k$-dimensional cost vector to each state.
\end{defi}
\vspace{-3mm}
\subsection{Supervisory Control and Scalability}
\vspace{-3mm}
Supervisory control theory (SCT) guarantees safe DES operation by modeling the plant $G$ with event set $\Sigma = \Sigma_c \cup \Sigma_{uc}$ (controllable and uncontrollable events). The plant is composed with a specification $E$ to obtain the desired behavior $K = \mathcal{L}_m(G \parallel E)$. If enforcing $K$ would disable uncontrollable events, the supervisor instead implements the supremal controllable and nonblocking sublanguage $\text{Sup}\,\mathcal{C}(K, G)$.

For large multi-agent systems, monolithic supervisor synthesis becomes computationally prohibitive because the state space grows combinatorially with the number of agents. This work adopts the scalable approach of \cite{liu2019scalable}, which exploits structural symmetry among groups of identical agents. A relabeling map $R: \Sigma \to T$ abstracts each group $\mathcal{G}_i$ of $n_i$ identical agents into a single template generator $\mathbf{H}_i$. Supervisory synthesis is performed on these template generators rather than on the full synchronous product. Under standard assumptions (disjoint agent event sets, relabeling-invariant specification, etc.), the resulting supervisor is nonblocking, and its state size and synthesis cost are independent of the agent population sizes $n_i$.

\vspace{-3mm}
\subsection{\GR{Automata-Based Prediction Model}}
\vspace{-3mm}
{\GR{Model predictive control is founded on the integration of optimization techniques with control theory, wherein} a finite-horizon \GR{optimal control problem is solved} at each \GR{time} step to enforce \GR{system} constraints and minimize a \GR{defined} cost \GR{functional} \citep{camacho2007model}.} 

{\GR{In this work, we propose a model predictive supervisory control which extends this concept} to DES by using the supervisor’s behavior as the predictive model in the receding-horizon scheme. To support this formulation, a matrix encoding for graphs and automata is adopted, inspired by exact representations such as MILP-based DES scheduling \citep{kobetski2006scheduling}, state-equation modeling \citep{kobayashi2012deterministic}, and classical determinant-based graph theory \citep{harary1962determinant}. A DFA with $n$ states and $m$ events is then represented by three binary matrices capturing its transition structure}\GR{:} 
\begin{itemize}
\item Adjacency matrix $A \in \{0,1\}^{n \times n}$: $A_{i,j} = 1$ if state $q_j$ is reachable from $q_i$ by any event\GR{;}
\item Event reachability matrix $B \in \{0,1\}^{m \times n}$: $B_{e,j} = 1$ if state $q_j$ can be reached by event $\sigma_e$ from any state\GR{;}
\item Event availability matrix $C \in \{0,1\}^{n \times m}$: $C_{i,e} = 1$ if event $\sigma_e$ is enabled at state $q_i$.
\end{itemize}

States and events are represented using one-hot encoding: $x \in \{0,1\}^n$ for states and $u \in \{0,1\}^m$ for events, where exactly one component is active ($\mathbf{1}^\top x = 1$ and $\mathbf{1}^\top u = 1$). Given the current state $x_t \in \{0,1\}^n$ and the applied event $u_t \in \{0,1\}^m$ at time $t$, the DFA dynamics are captured algebraically by the state transition model
\begin{equation}
x_{t+1} = \big( A\,x_t \big) \odot \big( B\,u_t \big),
\label{eq:dinamica}
\end{equation}
where $\odot$ denotes  Hadamard (element-wise) multiplication     \citep{million2007hadamard}. The resulting vector $x_{t+1}$ is the unique state satisfying both reachability constraints simultaneously. The correctness of this algebraic transition relies on a property formalized as follows. The matrices $A, B, C$ represent a deterministic automaton $G$ if each state-event pair $(x_t, u_t)$ yields at most one successor state. Algebraically, this condition is expressed as $
\big(A \cdot B^\top \big) \odot C = C.
\label{defi:requisito}$
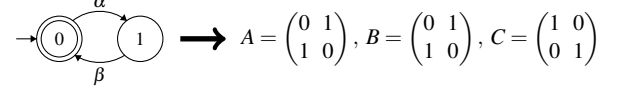
\begin{figure}[htbp]
    \centering
    \vspace{-2mm}
    \begin{tikzpicture}[scale=0.1]
        \scriptsize
        \tikzstyle{every node}+=[inner sep=0pt]

        % =========================
        % Automaton G
        % =========================
        \draw [black] (16,-29.2) circle (3);
        \draw (16,-29.2) node {$0$};
        \draw [black] (16,-29.2) circle (2.4);

        \draw [black] (26.7,-29.2) circle (3);
        \draw (26.7,-29.2) node {$1$};

        \draw [black] (10.3,-29.2) -- (13,-29.2);
        \fill [black] (13,-29.2) -- (12.2,-28.7) -- (12.2,-29.7);

        \draw [black] (17.765,-26.816) arc (128.53908:51.46092:5.754);
        \fill [black] (24.93,-26.82) -- (24.62,-25.93) -- (24,-26.71);
        \draw (21.35,-25.06) node [above] {$\alpha$};

        \draw [black] (24.659,-31.357) arc (-57.40945:-122.59055:6.142);
        \fill [black] (18.04,-31.36) -- (18.45,-32.21) -- (18.98,-31.37);
        \draw (21.35,-32.82) node [below] {$\beta$};

        % =========================
        % Arrow from automaton to matrices
        % =========================
\draw[->, line width=2pt] (32,-29.2) -- (38,-29.2);
        % =========================
        % Matrices
        % =========================
\node[anchor=west] at (40,-29.2) {%
\scalebox{1.2}{$%
\displaystyle
A =
\begin{pmatrix}
0 & 1 \\
1 & 0
\end{pmatrix},
\;
B =
\begin{pmatrix}
0 & 1 \\
1 & 0
\end{pmatrix},
\;
C =
\begin{pmatrix}
1 & 0 \\
0 & 1
\end{pmatrix}
$}%
};
    \end{tikzpicture}
    \vspace{-3mm}
    \caption{ Example \ref{ex:example1}: automaton $G$ and the matrices $A$, $B$, and $C$. \vspace{-2mm}}
    \label{fig:example_automaton}
\end{figure}
\begin{myexmp}\label{ex:example1}
Consider the automaton in Fig.~\ref{fig:example_automaton}. Its algebraic representation follows \eqref{eq:dinamica} with matrices $A$, $B$, $C$ given in \eqref{eq:example1matrix}. Applying \eqref{eq:dinamica} with $x_t=(1\quad 0)^T$ and $u_t=(1\quad 0)^T$yields
\begin{small}
\begin{equation}
x_{t+1} =
\begin{pmatrix}
0 & 1 \\
1 & 0
\end{pmatrix}\,\begin{pmatrix}
 1 \\ 0
\end{pmatrix}\odot \begin{pmatrix}
0 & 1 \\
1 & 0
\end{pmatrix}\,\begin{pmatrix}
 1 \\ 0
\end{pmatrix}= \begin{pmatrix}
 0 \\ 1
\end{pmatrix}\odot  \begin{pmatrix}
 0 \\ 1
\end{pmatrix}= \begin{pmatrix}
 0 \\ 1
\end{pmatrix}.
\end{equation}
\end{small}
\end{myexmp}

\vspace{-3mm}

%% file: sections_V2/03Problema.tex
\label{sec:problem}
\vspace{-3mm}
{This section formalizes the UTM problem as a hierarchical multi-agent coordination \GR{challenge} in a constrained urban airspace. The objective is to \GR{efficiently} coordinate a fleet of UAVs \GR{to successfully execute} assigned delivery tasks \GR{while simultaneously} respecting safety, capacity, and regulatory constraints. The \GR{resulting formulation is designed to prioritize} scalability, adaptability, and operational compliance in dense environments. The urban airspace setting is \GR{characterized by} three entities:}
\begin{itemize}
    \item UAV Fleet $\mathcal{U}$: A set of \GR{UAVs, where} $i \in \mathcal{U}$\GR{, are tasked with} performing missions. Each UAV is modeled as a weighted automaton (Def.~\ref{def:weighted})
     \begin{equation}
     G_i = (Q_i, \Sigma_i, \delta_i, q_{0_i}, Q_{m_i}, W_i),
     \end{equation}
      {where $W_i$ assigns energy, time, and penalty weights used later by the MPSC optimization \GR{process};}
    \item Multilayer Airspace: the environment is represented as a directed multigraph. Vertices comprise special nodes (vertiports, suppliers, clients, charging stations) and logical nodes (intermediate routing points). Logical nodes carry a physical position $(x,y,z)$ and may share $(x,y)$ while differing in altitude, representing different layers. Each airway  is modeled as a non-directed edge whose operational direction is assigned at runtime \citep{labib2019multilayer}.
    \item UTM: A global discrete-event controller responsible for enforcing shared-resource and regulatory constraints. It \GR{actively} prevents edge/vertex conflicts, maintains online geofences, and regulates fleet-level access to the airspace through a set of supervisory automata.
\end{itemize}

The multi-agent coordination problem requires event sequences that complete all missions subject to battery, workflow, communication, and global airspace safety constraints. A hierarchical solution separates the UAV layer, where agents execute controllable events (edge acquisition, task execution, charging) based on their weighted automata, from the UTM layer, which tracks the fleet configuration, computes prohibited transitions, and enforces geofencing and mutex constraints on shared vertices and edges. A bidirectional interface enables UAVs to send edge requests and state updates while the UTM returns disabled-event sets, yielding scalable, safe, regulation-compliant multi-UAV operation.

%% file: sections_V2/04Metodologia.tex
\label{sec:methodology}
\vspace{-3mm}
\GR{Focusing on large multi-agent systems, this section presents a hierarchical and scalable model predictive supervisory control design for coordination. The framework achieves a comprehensive mathematical formulation by integrating receding-horizon optimal control problem with scalable solution methods.}
\vspace{-3mm}
\subsection{Multilayer Airspace Modeling}
\label{subsec:airspace_modeling}
\vspace{-3mm}
The urban low-altitude airspace is modeled as a multilayer multigraph $\mathcal{M}=(\mathcal{V},\mathcal{A})$ whose layers $\ell\in\mathcal{L}$ represent altitude bands \citep{labib2019multilayer}. Nodes are either special $\mathcal{V}_S$ (mission locations) or logical $\mathcal{V}_L$ (navigation waypoints, each with horizontal coordinates and a layer; distinct logical nodes may share horizontal coordinates across layers). Undirected corridors connect nodes within or between layers; special nodes are linked only through logical nodes. The admissible corridor set $\mathcal{A} = \mathcal{A}_{SL} \cup \mathcal{A}_{LL} \cup \mathcal{A}_V$ consists of special-to-logical corridors, logical-to-logical corridors that replicate a base-layer pattern, and vertical corridors connecting logical nodes with the same horizontal coordinates in different layers. No two corridors intersect except at shared endpoints.

\begin{figure*}[t]
\centering
\includegraphics[width=0.65\linewidth, trim={1.4cm 1cm 1.5cm 0.85cm}, clip]{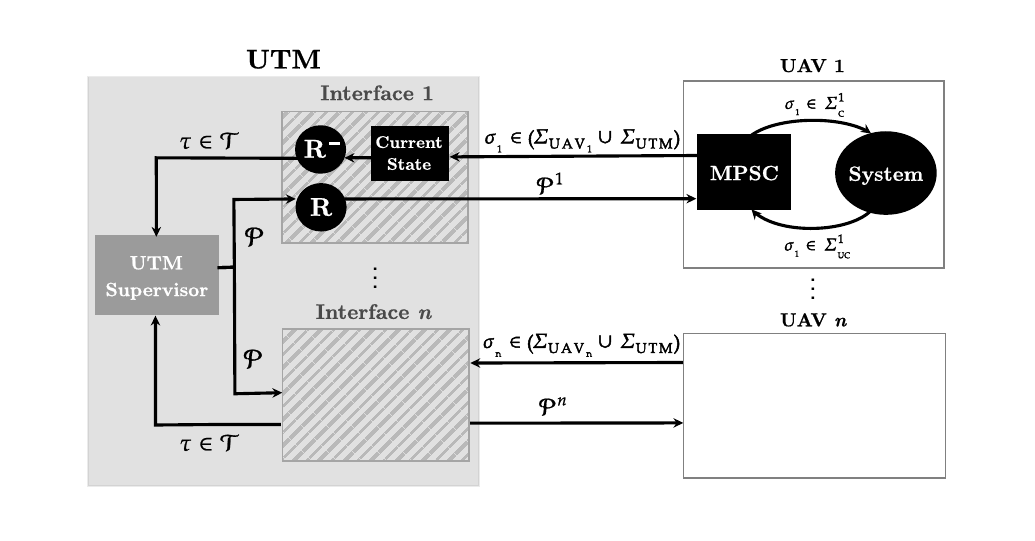}
\vspace{-3mm}
\caption{Hierarchical and scalable supervisory architecture.} 
\label{fig:hierarchical_utm_architecture}
\end{figure*}

\begin{myexmp}
\label{ex:minimal-scenario}
A minimal delivery scenario \GR{is used to} illustrate the framework. The environment contains the  special nodes $\mathcal{V}_S=\{V, S, C, E\}$, \GR{which represent} a vertiport ($V$), a supplier ($S$), a client ($C$), and a charging station ($E$)\GR{, respectively. These} are connected through a single  logical  node $\mathcal{V}_L=\{L\}$, \GR{considering} only a single layer ($L=1$). \GR{This} setting is shown in Fig.~\ref{fig:running_example_layout}.
\begin{figure}[htbp]
    \centering
    \includegraphics[width=0.35\columnwidth, trim={10cm 8.5cm 10cm 9.3cm}, clip]{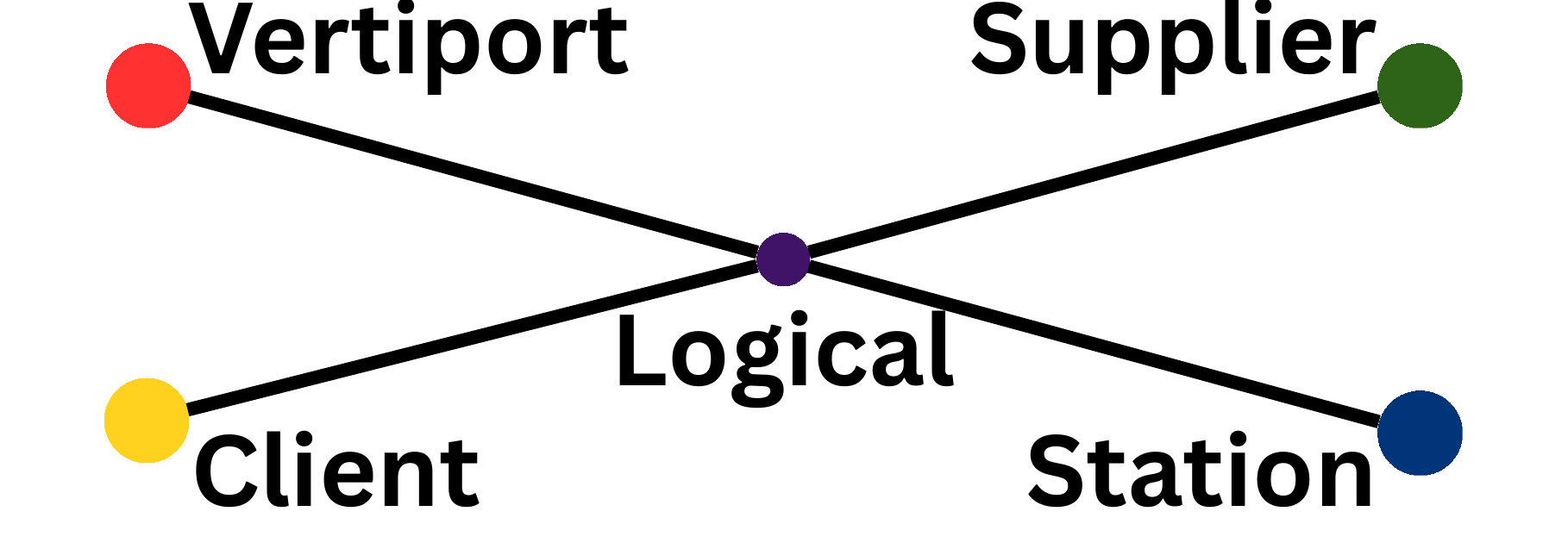}
    \caption{Single layer representation of the delivery scenario.}
    \label{fig:running_example_layout}
\end{figure}
\end{myexmp}
\vspace{-6mm}
\subsection{Hierarchical \& Scalable Supervisory Control Archi\-tec\-tu\-re}%} Supervisory Control Solution }
\vspace{-3mm}
The proposed hierarchical MPSC framework couples SCT-based admissibility with MILP-based receding-horizon optimization, addressing the state-space explosion of multi-agent supervisory control. The UTM supervisory layer enforces global admissibility by modeling shared airspace as a DES resource-allocation problem. A class-level supervisor template, a DFA over a generic alphabet encoding the admissible behavior shared by all UAVs of the same class, is synthesized once for a representative UAV model and instantiated for each vehicle via event relabeling \citep{liu2019scalable,cai2010supervisor,dulce2022distributed}, yielding indexed instances with vehicle-specific event labels. Using relabeled events, the UTM maintains a global view of vertex occupancy, corridor availability, directional flow, vertical access, and dynamic geofences, and computes a prohibited-event set $\mathcal{P}$ that disables movements violating mutual exclusion, geofencing, direction rules, or shared-resource constraints.

The UAV decision layer receives $\mathcal{P}$ and solves a finite-horizon MPSC problem within its relabeled supervisor's reachable sub-automaton, selecting a cost-minimizing sequence over mission progress, travel, energy, waiting, charging, and return-to-base behavior. Only the first event is executed, after which the optimization is repeated from the updated state in a receding-horizon manner. This separation defines the hierarchy: each UAV makes autonomous, locally cost-optimal decisions strictly within the admissible set enforced by the UTM. SCT guarantees language-level safety and nonblockingness, while MPSC supplies online cost optimization. Under synchronized state updates, every executed event satisfies both the supervised language and current global restrictions.

Assuming that the supervisors are controllable and nonblocking, $\mathcal{P}$ is synchronized across UAVs, and each UAV executes only events permitted by its supervisor and not in $\mathcal{P}$, every executed sequence remains in the admissible supervised language and respects the UTM's mutual-exclusion and geofencing constraints.

This property follows from the fact that the MPSC optimization is restricted to the event set admitted by the SCT supervisor after applying the prohibitions imposed by the UTM coordination layer. Therefore, optimization affects only the selection among admissible behaviors and does not alter the supervisory safety constraints.
\vspace{-3mm}
\subsection{UAV DES Models and Formal Specifications}
\vspace{-3mm}
The UAV decision-making framework derives its alphabet directly from the multilayer airspace graph: each directed edge yields acquisition and release symbols for movement rights, while additional symbols capture task actions (pickup, delivery, charging), as well as global functions such as task acceptance, mission completion, and battery reporting. From this alphabet, plant automata are built to reflect operational capabilities: a movement automaton models stationary/transit alternation, edge-resource plants regulate traversal direction, an operational-mode plant encodes mission activities (servicing, delivery, charging), and auxiliary automata handle communication, liveness, and battery monitoring. Because the DES structure originates from the airspace graph, the number of events and automata grows with the vertices, corridors, layers, and resources; each corridor adds acquisition and release events, and each additional layer expands admissible movement and edge constraints.

\begin{figure}[!htbp]
\centering
\scriptsize

\begin{tabular}{cc}
\begin{minipage}{0.6\linewidth}
\begin{minipage}{0.7\linewidth}
\centering
\subfloat[$G_{\mathrm{move}}$\label{fig:move_automaton}]{
\begin{tikzpicture}[scale=0.11]
\tikzstyle{every node}+=[inner sep=0pt]
\draw [black] (17.3,-22.8) circle (3);
\draw (17.3,-22.8) node {$q_{\mathrm{idle}}$};
\draw [black] (17.3,-22.8) circle (2.4);
\draw [black] (59.8,-22.8) circle (3);
\draw (59.8,-22.8) node {$q_{\mathrm{mov}}$};
\draw [black] (10.9,-22.8) -- (14.3,-22.8);
\fill [black] (14.3,-22.8) -- (13.5,-22.3) -- (13.5,-23.3);
\draw [black] (20.112,-21.755) arc (108.87893:71.12107:56.984);
\fill [black] (56.99,-21.76) -- (56.39,-21.02) -- (56.07,-21.97);
\draw (38.55,-18.19) node [above] {$t_{VL},t_{LV},t_{SL},t_{LS},t_{CL},t_{LC},t_{EL},t_{LE}$};
\draw [black] (56.916,-23.624) arc (-75.24026:-104.75974:72.089);
\fill [black] (20.18,-23.62) -- (20.83,-24.31) -- (21.09,-23.34);
\draw (38.55,-26.5) node [below] {$r_{VL},r_{LV},r_{SL},r_{LS},r_{CL},r_{LC},r_{EL},r_{LE}$};
\end{tikzpicture}
}
\end{minipage}
\begin{minipage}{0.45\linewidth}
\centering
\subfloat[$G_{\mathrm{edge,dir}}$\label{fig:edge_dir_automaton}]{
\begin{tikzpicture}[scale=0.11]
\tikzstyle{every node}+=[inner sep=0pt]

% States
\draw [black] (13.4,-17.2) circle (3);
\draw (13.4,-17.2) node {$q_{\mathrm{free}}$};
\draw [black] (13.4,-17.2) circle (2.4);

\draw [black] (22.7,-7.3) circle (3);
\draw (22.7,-7.3) node {$q_{\mathrm{occ^+}}$};

\draw [black] (23.3,-26.2) circle (3);
\draw (23.3,-26.2) node {$q_{\mathrm{occ^-}}$};

% Initial arrow
\draw [black] (7.4,-17.2) -- (10.4,-17.2);
\fill [black] (10.4,-17.2) -- (9.6,-16.7) -- (9.6,-17.7);

% Forward transitions
\draw [black] (14.571,-14.443) arc (151.50918:122.07064:15.666);
\fill [black] (20.02,-8.64) -- (19.08,-8.64) -- (19.61,-9.49);
\draw (16.39,-9.72) node [left] {$t_{uv}$};

\draw [black] (21.082,-9.824) arc (-35.61924:-50.80094:29.096);
\fill [black] (15.82,-15.43) -- (16.75,-15.31) -- (16.12,-14.53);
\draw (21,-12) node [right] {$r_{uv}$};

% Reverse transitions
\draw [black] (20.558,-24.992) arc (-118.93092:-145.61645:16.676);
\fill [black] (20.56,-24.99) -- (20.1,-24.17) -- (19.62,-25.04);
\draw (15.18,-23.22) node [below] {$t_{vu}$};

\draw [black] (16.075,-18.553) arc (58.82193:36.63069:19.75);
\fill [black] (16.07,-18.55) -- (16.5,-19.39) -- (17.02,-18.54);
\draw (22,-22.5) node [above] {$r_{vu}$};

\end{tikzpicture}
}
\end{minipage}
\begin{minipage}{0.45\linewidth}
\centering
\subfloat[$G_{\mathrm{modes}}$\label{fig:task_automaton}]{
\begin{tikzpicture}[scale=0.11]
\tikzstyle{every node}+=[inner sep=0pt]

% States
\draw [black] (13.4,-17.2) circle (3);
\draw (13.4,-17.2) node {$q_{\mathrm{base}}$};
\draw [black] (13.4,-17.2) circle (2.4);

\draw [black] (29.3,-17.2) circle (3);
\draw (29.3,-17.2) node {$q_{\mathrm{place}}$};

\draw [black] (22.7,-7.3) circle (3);
\draw (22.7,-7.3) node {$q_{\mathrm{pick}}$};

\draw [black] (23.3,-26.2) circle (3);
\draw (23.3,-26.2) node {$q_{\mathrm{load}}$};

% Initial arrow
\draw [black] (7.4,-17.2) -- (10.4,-17.2);
\fill [black] (10.4,-17.2) -- (9.6,-16.7) -- (9.6,-17.7);

% Supplier transitions
\draw [black] (14.571,-14.443) arc (151.50918:122.07064:15.666);
\fill [black] (20.02,-8.64) -- (19.08,-8.64) -- (19.61,-9.49);
\draw (16.39,-9.72) node [left] {$sw_S$};

\draw [black] (21.082,-9.824) arc (-35.61924:-50.80094:29.096);
\fill [black] (15.82,-15.43) -- (16.75,-15.31) -- (16.12,-14.53);
\draw (21,-12) node [right] {$ew_S$};

% Client transitions
\draw [black] (16.323,-16.532) arc (99.92622:80.07378:29.161);
\fill [black] (26.38,-16.53) -- (25.67,-15.9) -- (25.5,-16.89);
\draw (21.35,-15.6) node [above] {$sw_C$};

\draw [black] (26.331,-17.629) arc (-83.67507:-96.32493:45.217);
\fill [black] (16.37,-17.63) -- (17.11,-18.21) -- (17.22,-17.22);
\draw (21.35,-18.4) node [below] {$ew_C$};

% Charging transitions
\draw [black] (20.558,-24.992) arc (-118.93092:-145.61645:16.676);
\fill [black] (20.56,-24.99) -- (20.1,-24.17) -- (19.62,-25.04);
\draw (15.18,-23.22) node [below] {$sc_E$};

\draw [black] (16.075,-18.553) arc (58.82193:36.63069:19.75);
\fill [black] (16.07,-18.55) -- (16.5,-19.39) -- (17.02,-18.54);
\draw (22,-22.5) node [above] {$ec_E$};

\end{tikzpicture}
}
\end{minipage}
\end{minipage}
&
\begin{minipage}{0.2\linewidth}
\centering

\subfloat[$G_{\mathrm{com}}$]{
\begin{tikzpicture}[scale=0.11]
\scriptsize
\tikzstyle{every node}+=[inner sep=0pt]
\draw [black] (30.9,-20.4) circle (3);
\draw [black] (30.9,-20.4) circle (2.4);
\draw (30.9,-20.4) node {$q_{\mathrm{com}}$};

\draw [black] (24.8,-20.4) -- (27.9,-20.4);
\fill [black] (27.9,-20.4) -- (27.1,-19.9) -- (27.1,-20.9);

\draw [black] (29.577,-17.72) arc (234:-54:2.25);
\draw (30.9,-13.15) node [above] {$ac,ft$};
\fill [black] (32.22,-17.72) -- (33.1,-17.37) -- (32.29,-16.78);
\end{tikzpicture}
}

\subfloat[$G_{\mathrm{live}}$]{
\begin{tikzpicture}[scale=0.11]
\scriptsize
\tikzstyle{every node}+=[inner sep=0pt]
\draw [black] (30.9,-20.4) circle (3);
\draw [black] (30.9,-20.4) circle (2.4);
\draw (30.9,-20.4) node {$q_{\mathrm{live}}$};

\draw [black] (24.8,-20.4) -- (27.9,-20.4);
\fill [black] (27.9,-20.4) -- (27.1,-19.9) -- (27.1,-20.9);

\draw [black] (29.577,-17.72) arc (234:-54:2.25);
\draw (30.9,-13.15) node [above] {$hb$};
\fill [black] (32.22,-17.72) -- (33.1,-17.37) -- (32.29,-16.78);
\end{tikzpicture}
}

\subfloat[$G_{\mathrm{bat}}$]{
\begin{tikzpicture}[scale=0.11]
\scriptsize
\tikzstyle{every node}+=[inner sep=0pt]
\draw [black] (30.9,-20.4) circle (3);
\draw [black] (30.9,-20.4) circle (2.4);
\draw (30.9,-20.4) node {$q_{\mathrm{pwr}}$};

\draw [black] (24.8,-20.4) -- (27.9,-20.4);
\fill [black] (27.9,-20.4) -- (27.1,-19.9) -- (27.1,-20.9);

\draw [black] (29.577,-17.72) arc (234:-54:2.25);
\draw (30.9,-13.15) node [above] {$lb$};
\fill [black] (32.22,-17.72) -- (33.1,-17.37) -- (32.29,-16.78);
\end{tikzpicture}
}

\end{minipage}
\end{tabular}
\vspace{-3mm}
\caption{UAV subplants}
\label{fig:uav_automata_new}
\end{figure}
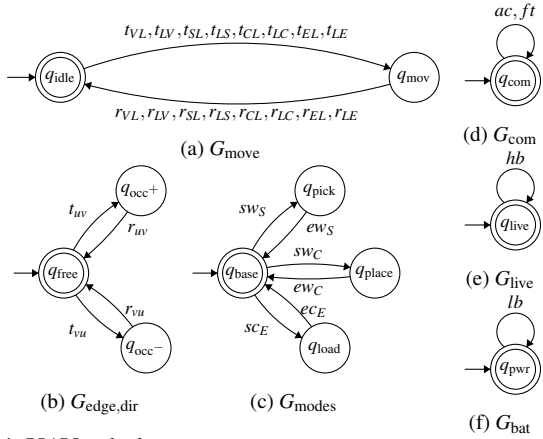

\begin{myexmp}
\label{ex:platas}
For the system model in Example~\ref{ex:minimal-scenario}, the UAV automata (Fig.~\ref{fig:uav_automata_new}) operate over alphabet $\Sigma_{\mathrm{UAV}}$ containing corridor acquisition and release events $t_{uv}$, $r_{uv}$; service events $sw_S, ew_S, sw_C, ew_C$; charging events $sc_E, ec_E$; and auxiliary events $ac, ft, hb, lb$ for task acceptance, heartbeat, and battery monitoring.
\end{myexmp}

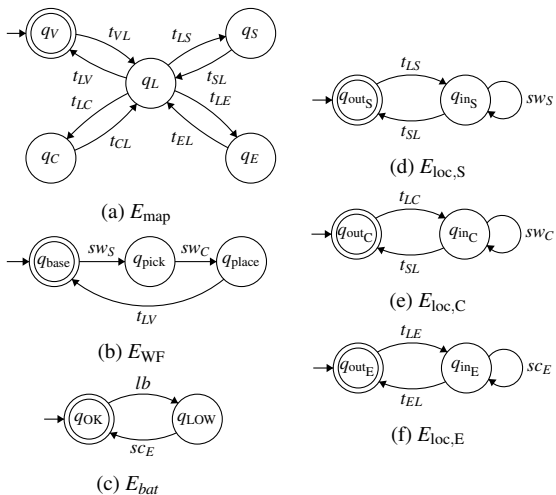
\begin{figure}[!htbp]
\centering
\scriptsize
\begin{tabular}{cc}
\begin{minipage}{0.4\linewidth}
\centering
% --- (a) Map specification ---
\subfloat[$E_{\mathrm{map}}$]{
\label{fig:emap}
\begin{tikzpicture}[scale=0.11]
\tikzstyle{every node}+=[inner sep=0pt]
\draw [black] (14.9,-13.7) circle (3);
\draw (14.9,-13.7) node {$q_V$};
\draw [black] (14.9,-13.7) circle (2.4);
\draw [black] (38.9,-13.7) circle (3);
\draw (38.9,-13.7) node {$q_S$};
\draw [black] (26.9,-19.7) circle (3);
\draw (26.9,-19.7) node {$q_L$};
\draw [black] (14.9,-28.7) circle (3);
\draw (14.9,-28.7) node {$q_C$};
\draw [black] (38.9,-28.7) circle (3);
\draw (38.9,-28.7) node {$q_E$};
\draw [black] (9.6,-13.7) -- (11.9,-13.7);
\fill [black] (11.9,-13.7) -- (11.1,-13.2) -- (11.1,-14.2);
\draw [black] (17.891,-13.816) arc (81.19277:45.67713:13.052);
\fill [black] (25.01,-17.38) -- (24.79,-16.46) -- (24.09,-17.18);
\draw (23.25,-14.53) node [above] {$t_{VL}$};
\draw [black] (23.971,-19.062) arc (-106.31478:-126.81532:21.372);
\fill [black] (17.17,-15.66) -- (17.51,-16.54) -- (18.11,-15.74);
\draw (18.89,-18.17) node [below] {$t_{LV}$};
\draw [black] (16.822,-26.399) arc (136.87202:116.86777:26.379);
\fill [black] (16.82,-26.4) -- (17.73,-26.16) -- (17,-25.47);
\draw (18.71,-22.83) node [above] {$t_{LC}$};
\draw [black] (25.198,-22.167) arc (-39.08336:-67.17685:19.193);
\fill [black] (25.2,-22.17) -- (24.31,-22.47) -- (25.08,-23.1);
\draw (23.36,-25.92) node [below] {$t_{CL}$};
\draw [black] (28.975,-17.539) arc (130.74572:102.38438:15.863);
\fill [black] (35.93,-14.06) -- (35.04,-13.75) -- (35.25,-14.72);
\draw (30.71,-14.86) node [above] {$t_{LS}$};
\draw [black] (36.63,-15.658) arc (-53.22064:-73.64926:21.443);
\fill [black] (29.83,-19.06) -- (30.74,-19.31) -- (30.46,-18.35);
\draw (34.9,-18.17) node [below] {$t_{SL}$};
\draw [black] (29.763,-20.585) arc (68.07757:38.18264:18.139);
\fill [black] (37.25,-26.2) -- (37.15,-25.26) -- (36.36,-25.88);
\draw (35.42,-22.4) node [above] {$t_{LE}$};
\draw [black] (36.174,-27.45) arc (-117.64499:-136.09481:28.504);
\fill [black] (28.86,-21.97) -- (29.06,-22.89) -- (29.78,-22.2);
\draw (30.76,-25.5) node [below] {$t_{EL}$};
\end{tikzpicture}
}
\vspace{0.1cm}
% --- (b or f) workflow / task specification ---
\subfloat[$E_{\mathrm{WF}}$]{
\label{fig:eworkflow}
\begin{tikzpicture}[scale=0.11]
\tikzstyle{every node}+=[inner sep=0pt]
\draw [black] (32.8,-22.6) circle (3);
\draw (32.8,-22.6) node {$q_{\mathrm{pick}}$};
\draw [black] (43.8,-22.6) circle (3);
\draw (43.8,-22.6) node {$q_{\mathrm{place}}$};
\draw [black] (21.3,-22.6) circle (3);
\draw [black] (21.3,-22.6) circle (2.4);
\draw (21.3,-22.6) node {$q_{\mathrm{base}}$};
\draw [black] (15.6,-22.6) -- (18.3,-22.6);
\fill [black] (18.3,-22.6) -- (17.5,-22.1) -- (17.5,-23.1);
\draw [black] (24.3,-22.6) -- (29.8,-22.6);
\fill [black] (29.8,-22.6) -- (29,-22.1) -- (29,-23.1);
\draw (27.05,-22) node [above] {$sw_S$};
\draw [black] (41.679,-24.714) arc (-51.02586:-128.97414:14.514);
\fill [black] (23.42,-24.71) -- (23.73,-25.61) -- (24.36,-24.83);
\draw (32.55,-28.44) node [below] {$t_{LV}$};
\draw [black] (35.8,-22.6) -- (40.8,-22.6);
\fill [black] (40.8,-22.6) -- (40,-22.1) -- (40,-23.1);
\draw (38.3,-22) node [above] {$sw_C$};
\end{tikzpicture}
}

\vspace{0.1cm}
% --- Battery coordination ---
\subfloat[$E_{bat}$]{
\label{fig:ebat}
\begin{tikzpicture}[scale=0.11]
\tikzstyle{every node}+=[inner sep=0pt]
\draw [black] (21.5,-23.1) circle (3);
\draw (21.5,-23.1) node {$q_{\mathrm{OK}}$};
\draw [black] (21.5,-23.1) circle (2.4);
\draw [black] (34.5,-23.1) circle (3);
\draw (34.5,-23.1) node {$q_{\mathrm{LOW}}$};
\draw [black] (16.1,-23.1) -- (18.5,-23.1);
\fill [black] (18.5,-23.1) -- (17.7,-22.6) -- (17.7,-23.6);
\draw [black] (23.796,-21.193) arc (119.61375:60.38625:8.508);
\fill [black] (32.2,-21.19) -- (31.76,-20.36) -- (31.26,-21.23);
\draw (28,-19.58) node [above] {$lb$};
\draw [black] (32.004,-24.743) arc (-65.53789:-114.46211:9.67);
\fill [black] (24,-24.74) -- (24.52,-25.53) -- (24.93,-24.62);
\draw (28,-26.11) node [below] {$sc_E$};
\end{tikzpicture}
}

\end{minipage}
&

\begin{minipage}{0.4\linewidth}
\centering

% --- Location-task S ---
\subfloat[$E_{\mathrm{loc,S}}$]{
\label{fig:elocS}
\begin{tikzpicture}[scale=0.11]
\tikzstyle{every node}+=[inner sep=0pt]
\draw [black] (21.5,-23.1) circle (3);
\draw (21.5,-23.1) node {$q_{\mathrm{out_S}}$};
\draw [black] (21.5,-23.1) circle (2.4);
\draw [black] (34.5,-23.1) circle (3);
\draw (34.5,-23.1) node {$q_{\mathrm{in_S}}$};
\draw [black] (16.1,-23.1) -- (18.5,-23.1);
\fill [black] (18.5,-23.1) -- (17.7,-22.6) -- (17.7,-23.6);
\draw [black] (23.796,-21.193) arc (119.61375:60.38625:8.508);
\fill [black] (32.2,-21.19) -- (31.76,-20.36) -- (31.26,-21.23);
\draw (28,-19.58) node [above] {$t_{LS}$};
\draw [black] (32.004,-24.743) arc (-65.53789:-114.46211:9.67);
\fill [black] (24,-24.74) -- (24.52,-25.53) -- (24.93,-24.62);
\draw (28,-26.11) node [below] {$t_{SL}$};
\draw [black] (37.18,-21.777) arc (144:-144:2.25);
\draw (41.75,-23.1) node [right] {$sw_S$};
\fill [black] (37.18,-24.42) -- (37.53,-25.3) -- (38.12,-24.49);
\end{tikzpicture}
}

\vspace{0.1cm}

% --- Location-task C ---
\subfloat[$E_{\mathrm{loc,C}}$]{
\label{fig:elocC}
\begin{tikzpicture}[scale=0.11]
\tikzstyle{every node}+=[inner sep=0pt]
\draw [black] (21.5,-23.1) circle (3);
\draw (21.5,-23.1) node {$q_{\mathrm{out_C}}$};
\draw [black] (21.5,-23.1) circle (2.4);
\draw [black] (34.5,-23.1) circle (3);
\draw (34.5,-23.1) node {$q_{\mathrm{in_C}}$};
\draw [black] (16.1,-23.1) -- (18.5,-23.1);
\fill [black] (18.5,-23.1) -- (17.7,-22.6) -- (17.7,-23.6);
\draw [black] (23.796,-21.193) arc (119.61375:60.38625:8.508);
\fill [black] (32.2,-21.19) -- (31.76,-20.36) -- (31.26,-21.23);
\draw (28,-19.58) node [above] {$t_{LC}$};
\draw [black] (32.004,-24.743) arc (-65.53789:-114.46211:9.67);
\fill [black] (24,-24.74) -- (24.52,-25.53) -- (24.93,-24.62);
\draw (28,-26.11) node [below] {$t_{SL}$};
\draw [black] (37.18,-21.777) arc (144:-144:2.25);
\draw (41.75,-23.1) node [right] {$sw_C$};
\fill [black] (37.18,-24.42) -- (37.53,-25.3) -- (38.12,-24.49);
\end{tikzpicture}
}

\vspace{0.1cm}

% --- Location-task E ---
\subfloat[$E_{\mathrm{loc,E}}$]{
\label{fig:elocE}
\begin{tikzpicture}[scale=0.11]
\tikzstyle{every node}+=[inner sep=0pt]
\draw [black] (21.5,-23.1) circle (3);
\draw (21.5,-23.1) node {$q_{\mathrm{out_E}}$};
\draw [black] (21.5,-23.1) circle (2.4);
\draw [black] (34.5,-23.1) circle (3);
\draw (34.5,-23.1) node {$q_{\mathrm{in_E}}$};
\draw [black] (16.1,-23.1) -- (18.5,-23.1);
\fill [black] (18.5,-23.1) -- (17.7,-22.6) -- (17.7,-23.6);
\draw [black] (23.796,-21.193) arc (119.61375:60.38625:8.508);
\fill [black] (32.2,-21.19) -- (31.76,-20.36) -- (31.26,-21.23);
\draw (28,-19.58) node [above] {$t_{LE}$};
\draw [black] (32.004,-24.743) arc (-65.53789:-114.46211:9.67);
\fill [black] (24,-24.74) -- (24.52,-25.53) -- (24.93,-24.62);
\draw (28,-26.11) node [below] {$t_{EL}$};
\draw [black] (37.18,-21.777) arc (144:-144:2.25);
\draw (41.75,-23.1) node [right] {$sc_E$};
\fill [black] (37.18,-24.42) -- (37.53,-25.3) -- (38.12,-24.49);
\end{tikzpicture}
}

\end{minipage}
\end{tabular}
\vspace{-3mm}
\caption{UAV specifications}
\label{fig:uav_specifications}
\end{figure}

Formal specifications restrict UAV behavior to ensure safety, task correctness, and airspace compliance. The map-navigation specification (Fig.~\ref{fig:emap}) encodes multilayer topology by allowing only transitions along reachable corridors. Workflow specifications (Fig.~\ref{fig:eworkflow}) enforce the high-level mission sequence and prevent unsafe or premature transitions. Battery specifications (Fig.~\ref{fig:ebat}) impose energy-aware operation by restricting motion under low-energy conditions. Location–task specifications (Fig.~\ref{fig:elocS}, \ref{fig:elocC}, \ref{fig:elocE}) guarantee that service and charging actions occur exclusively when the UAV is at the corresponding functional node. Together, these components yield a complete DES formulation of the scenario.
   
Building upon these generic models, a monolithic supervisor $\mathcal{S}_{\mathrm{generic}}$ was synthesized for a single agent. This supervisor encodes the safe behavior of a UAV operating within the modeled logical network. To address the scalability challenges inherent in multi-agent discrete-event systems, this generic supervisor serves as a template that can be specialized for each UAV, $i \in \mathcal{U}$  by duplicating the automata and adding indices to the agent's events. 
\vspace{-3mm}
\subsection{UTM Models}
\label{sec:utm_models}
\vspace{-3mm}
The UTM layer is modeled as a scalable supervisor whose plants reuse the UAV's map and motion automata together with a vertex-blocking plant, while specifications enforce vertex blocking and edge mutual exclusion. The synthesized supervisor provides dynamic geofencing, corridor directionality enforcement, and shared-resource arbitration across the fleet. Its state-space size grows with the airspace graph (vertices, corridors, layers) but, due to relabeling and template instantiation, remains independent of the number of UAVs. New tasks are assigned to the first available, charged UAV.
\begin{myexmp}
\label{ex:utm} 
For the scenario of Example~\ref{ex:minimal-scenario}, the UTM alphabet is $\Sigma_{\mathrm{UTM}}$, containing all corridor events $t_{uv}$ and vertex-blocking events $b_u, ub_u$. Its plants are the map automaton (Fig.~\ref{fig:emap}), the directional edge-usage automaton (Fig.~\ref{fig:edge_dir_automaton}), and a geofence plant (Fig.~\ref{fig:block_automaton}) that generates blocking events. For each vertex, a specification $E_{\mathrm{block},i}$ (e.g., Fig.~\ref{fig:block_specifications}) disables incoming corridor events after the corresponding blocking event; a mutex specification $E_{\mathrm{mutex},i}$ (e.g., Fig.~\ref{fig:verticemutex}) ensures a vertex, once entered, remains occupied until a departure event occurs.
\vspace{-8mm}
\begin{figure}[!htbp]
\centering
\scriptsize
\begin{tabular}{ccc}
\subfloat[$G_{\mathrm{geofence}}$\label{fig:block_automaton}]{
\begin{tikzpicture}[scale=0.11]
\tikzstyle{every node}+=[inner sep=0pt]
\draw [black] (11,-19) circle (3);
\draw (11,-19) node {$q_{\mathrm{blk}}$};
\draw [black] (11,-19) circle (2.4);
\draw [black] (5.5,-19) -- (8,-19);
\fill [black] (8,-19) -- (7.2,-18.5) -- (7.2,-19.5);
\draw [black] (9.677,-16.32) arc (234:-54:2.25);
\draw (11,-11.75) node [above] {$\begin{matrix}
b_{S}, ub_{S}, b_{C}, ub_{C}, b_{E}, \\
ub_{E}, b_{L}, ub_{L},b_{V},ub_{V}
\end{matrix}$};
\fill [black] (12.32,-16.32) -- (13.2,-15.97) -- (12.39,-15.38);
\end{tikzpicture}
} 
&
\subfloat[$E_{\mathrm{block,L}}$\label{fig:block_specifications}]{
\begin{tikzpicture}[scale=0.11]
\tikzstyle{every node}+=[inner sep=0pt]
\draw [black] (11.1,-19) circle (3);
\draw (11.1,-19) node {$q_{\mathrm{ub_{L}}}$};
\draw [black] (11.1,-19) circle (2.4);
\draw [black] (23.4,-19) circle (3);
\draw (23.4,-19) node {$q_{\mathrm{b_{L}}}$};
\draw [black] (5.6,-19) -- (8.1,-19);
\fill [black] (8.1,-19) -- (7.3,-18.5) -- (7.3,-19.5);
\draw [black] (13.516,-17.247) arc (115.90899:64.09101:8.546);
\fill [black] (20.98,-17.25) -- (20.48,-16.45) -- (20.05,-17.35);
\draw (17.25,-15.89) node [above] {$b_{L}$};
\draw [black] (20.681,-20.248) arc (-72.76875:-107.23125:11.582);
\fill [black] (13.82,-20.25) -- (14.43,-20.96) -- (14.73,-20.01);
\draw (17.25,-21.27) node [below] {$ub_{L}$};
\draw [black] (9.777,-16.32) arc (234:-54:2.25);
\draw (11.1,-11.75) node [above] {$ t_{V L},t_{S L},t_{C L},t_{E L}$};
\fill [black] (12.42,-16.32) -- (13.3,-15.97) -- (12.49,-15.38);
\end{tikzpicture}

}
&
\subfloat[$E_{\mathrm{mutex,L}}$\label{fig:verticemutex}]{
\begin{tikzpicture}[scale=0.11]
\tikzstyle{every node}+=[inner sep=0pt]
\draw [black] (11.1,-19) circle (3);
\draw (11.1,-19) node {$q_{\mathrm{L_{free}}}$};
\draw [black] (11.1,-19) circle (2.4);
\draw [black] (23.4,-19) circle (3);
\draw (23.4,-19) node {$q_{\mathrm{L_{occ}}}$};
\draw [black] (5.6,-19) -- (8.1,-19);
\fill [black] (8.1,-19) -- (7.3,-18.5) -- (7.3,-19.5);
\draw [black] (13.516,-17.247) arc (115.90899:64.09101:8.546);
\fill [black] (20.98,-17.25) -- (20.48,-16.45) -- (20.05,-17.35);
\draw (17.25,-15.89) node [above] {$t_{V L}, t_{S L}, t_{C L}, t_{E L}$};
\draw [black] (20.681,-20.248) arc (-72.76875:-107.23125:11.582);
\fill [black] (13.82,-20.25) -- (14.43,-20.96) -- (14.73,-20.01);
\draw (17.25,-22.27) node [below] {$t_{L V}, t_{L S}, t_{L C}, t_{L E}$};
\end{tikzpicture}}
\end{tabular}
 \caption{Example \ref{ex:utm}: (a) Plant with vertex blocking related events; (b) Block vertex; %, $i\in \small \{V,S,C,E,L\}$; 
 (c) Mutual exclusion.}%, $i\in \small \{V,S,C,E,L\}$.}
\end{figure}
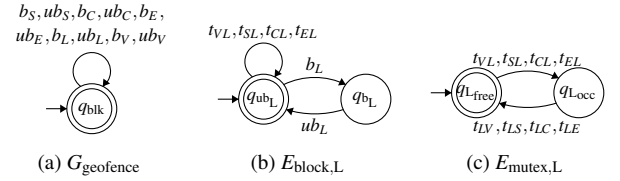
\end{myexmp}
\vspace{-10mm}
The UTM tracks each UAV's state within $\mathcal{S}_{\mathrm{UTM}}$ individually, avoiding an explicit synchronous product. Agent-specific events update only that UAV's state, while geofencing events $b_u$, $ub_u$ are applied to all states simultaneously. 

From the combined disabled transitions, the UTM computes the prohibited-event set $\mathcal{P}$ of all $t_{uv}$ that would violate mutex, geofencing, or shared-resource constraints, and broadcasts it so that each MPSC controller restricts its optimization to $\mathcal{P}$, ensuring decentralized, conflict-free coordination.
\vspace{-15mm}
\subsection{MILP Formulation for MPSC}
\vspace{-3mm}
When a new task is issued, it is assigned to a single UAV, which generates its corresponding desired event set $\mathcal{I}$. This set is composed of events related to the accomplishment of tasks. The UAV must solve an MPSC optimization problem to determine how to execute the task. 

Prior to optimization, each state of each plant and specification automaton of the UAV has a cost vector assigned; these atomic cost components are then algebraically lifted to the supervisor level \citep{loures2025adaptive}. A state-dependent sub-automaton is extracted around the UAV’s current supervisor state using a receding horizon of length $H$\GR{. On this structure}, a Mixed-Integer Linear Program is formulated. The MILP selects the next event by minimizing operational cost while maximizing incentives associated with the desired event set $\mathcal{I}$. The proposed MILP is posed as follows:
%\begin{small}
\begin{align}
\min \quad 
& \alpha_S\, \sum_{t=0}^{H-1}  \mathbf{w}^\top x_t 
 \;-\; \beta_I \sum_{i \in \mathcal{I}} \sum_{t=0}^{H-1} (H - t)\, u_{t,i}
 \label{obj_incentive_final} \\[1mm]
\text{s.t.} \quad 
& x_0 = q_0 
 \label{init_cond_final} \\
& x_{t+1} = (A x_t) \odot (B u_t) 
 && \forall t
 \label{dyn_new} \\
& \mathbf{1}_{|Q|}^\top x_t = 1 
 && \forall t
 \label{state_det_final} \\
& \mathbf{1}_{|\Sigma|}^\top u_t = 1 
 && \forall t
 \label{event_det_final} \\
& \mathbf{1}_{|\Sigma|}^\top \big( C x_t \odot u_t \big) = 1 
 && \forall t
 \label{event_feas_final} \\
& u_{t,i} = 0 
 && i \in \mathcal{P},\; \forall t
 \label{event_prohibit_final} \\
& x_t \in \{0,1\}^{|Q|},\quad u_t \in \{0,1\}^{|\Sigma|}
 && \forall t
 \label{bin_vars_final}
\end{align}
%\end{small}
\GR{with $H$ being the prediction horizon, $x_t \in \{0,1\}^{|Q|}$ the state vector, $u_t \in \{0,1\}^{|\Sigma|}$ the event vector, $q_0$ the initial state at time $t=0$, $\mathcal{I}$ the desired event set, $\mathcal{P}$ the set of prohibited events, $\mathbf{w}$ the aggregated operational cost in state $x_t$, and $\alpha_S$ and $\beta_I$ the optimization weights for operational cost and incentive term, respectively.}

%\begin{itemize}
 %\item $H$: The prediction horizon.
 %\item $x_t \in \{0,1\}^{|Q|}$: State vector and $u_t \in \{0,1\}^{|\Sigma|}$: Event vector.
 %\item $q_0$: The initial state vector at time $t=0$.
 %\item $\mathcal{I}$: The desired event set.
 %\item $\mathcal{P}$: The set of prohibited events.
 %\item $\mathbf{w}$: The aggregated operational cost in state $x_t$.
%\item \textbf{Optimization weights}: $\alpha_S$ for operational cost  and $\beta_I$ for  incentive term.
%\end{itemize}

%The MILP minimizes a multi-objective cost functional \eqref{obj_incentive_final}\GR{, which balances}  operational costs against and an incentive for desirable events. The solution is constrained by the DES algebraic framework: initial state fixation \eqref{init_cond_final}, automaton dynamics \eqref{dyn_new}, state and event one-hot encoding \eqref{state_det_final}--\eqref{event_det_final}, event feasibility via matrix $C$ \eqref{event_feas_final}, and supervisory prohibitions \eqref{event_prohibit_final}. This receding-horizon formulation operates cyclically: it first generates a depth-$H$ sub-automaton via BFS (Breadth-First Search) \citep{bundy1984breadth} and applies a structural correction step to ensure that the resulting sub-automaton satisfies the determinism requirement of Definition \ref{defi:requisito}. \PP{Then, the MILP problem is solved} using  the matrices \PP{$A$, $B$ and $C$} from the sub-automaton, executing only the first optimal event $u_0^\star$ updating to $x_{t+1}$ and repeating, and the entire procedure repeats at the next iteration. 

\GR{The MILP minimizes a multi-objective cost function \eqref{obj_incentive_final}, which balances operational costs against an incentive for desirable events. The solution is constrained by the DES algebraic framework, incorporating: initial state fixation \eqref{init_cond_final}, automaton dynamics \eqref{dyn_new}, state and event one-hot encoding \eqref{state_det_final}–\eqref{event_det_final}, event feasibility via matrix $C$ \eqref{event_feas_final}, and supervisory prohibitions \eqref{event_prohibit_final}.}

\GR{This receding-horizon formulation operates cyclically: it first generates a depth-$H$ sub-automaton via Breadth-First Search (BFS) \citep{bundy1984breadth}. It then applies a structural correction step to ensure the resulting sub-automaton satisfies the determinism requirement of Definition \ref{defi:requisito}. Subsequently, the MILP problem is solved using the matrices $A$, $B$, and $C$ from the sub-automaton, executing only the first optimal event $u_0^\star$, updating the state to $x_{t+1}$, and repeating the entire procedure at the next iteration. }

%% file: sections_V2/06Estudo.tex
\label{sec:results}
\vspace{-3mm}
To validate the proposed Hierarchical MPSC architecture, two experiments were executed within a ROS 2/Gazebo framework \citep{quigley2009ros}.\footnote{For access to the complete and functional code, please visit the project's GitHub repository: \url{https://github.com/lacsed/MPSC-Hierarchical-UTM.git}} The implementation employed a multi-process simulation architecture that leveraged the Python.NET bridge to execute the \textsc{UltraDES} .NET library \citep{alves2023ultrades} for supervisory control design, while the optimization layer utilized the Gurobi solver \citep{gurobi2022gurobi}. The case studies feature structured urban airspaces represented by planar H/V graphs with logical corridors and specialized service nodes. Scenario R1 contains 9 graph nodes, 15 air corridors, 1 vertiport, 1 charging station, 1 supplier, and 2 clients, executed with $N=2$ UAVs. Scenario R2 contains 19 graph nodes, 34 air corridors, 1 vertiport, 1 charging station, 2 suppliers, and 3 clients, executed with $N=4$ UAVs across two movement layers, and extends the running example (Examples~\ref{ex:example1}, \ref{ex:minimal-scenario}, \ref{ex:platas}, and \ref{ex:utm}) by augmenting the topology and fleet size.

For each scenario, the UAV model instantiates $G_{move}$, directional edge‑usage plants, $G_{modes}$, and the auxiliary plants $G_{com}$, $G_{live}$, and $G_{bat}$. Specifications comprise $E_{map}$, $E_{WF}$, $E_{bat}$, and location–task specifications $E_{loc,i}$ for the specialized nodes in $\mathcal{V}_S$.
\begin{figure}[htbp]
\centering

\begin{minipage}[c]{0.28\columnwidth}
    \centering
    \includegraphics[width=0.9\linewidth]{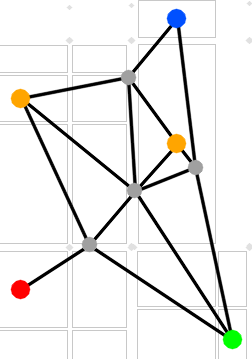}\\[-1mm]
    \small (a) R1
\end{minipage}
\hfill
\begin{minipage}[c]{0.28\columnwidth}
    \centering
    \includegraphics[width=0.9\linewidth]{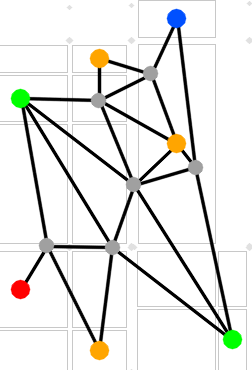}\\[-1mm]
    \small (b) R2
\end{minipage}
\hfill
\begin{minipage}[c]{0.38\columnwidth}
    \centering
    \scriptsize
    \begin{tabular}{|l|l|}
    \hline
        \textbf{Color} & \textbf{Meaning} \\
        \hline
        Gray & Logical node \\
        Red & Vertiport \\
        Blue & Charging station \\
        Green & Supplier \\
        Orange & Client \\
        \hline
    \end{tabular}
\end{minipage}

\vspace{-3mm}
\caption{Urban airspace topology for scenarios R1 and R2. \vspace{-3mm}}
\label{fig:airspace_topology}
\end{figure}

A monolithic supervisor is synthesized for the generic UAV model and is replicated through event relabeling. State‑cost vectors are computed offline after synthesis, each state receiving an atomic cost triple $(E,T_f,D)$ for energy, traversal time, and task incentives. The UTM model contains the global map, resource constraints, geofencing vertex‑blocking specifications, and vertex‑mutex specifications, as detailed in Section \ref{sec:utm_models}. Disabled UTM events are communicated to the fleet through the prohibited‑event set $\mathcal{P}$, whereas geofencing events $b_u$ and $ub_u$ are applied synchronously as they modify global admissibility.

\vspace{-1mm}
\begin{table}[htbp]
\centering
\scriptsize
\caption{DES metrics for R1.}\vspace{-2mm}
\label{tab:centralized_comparison}
\begin{tabular}{|l|c|c|}
\hline
\textbf{Metric} & \textbf{Hierarchical} & \textbf{Centralized exact} \\
\hline
Supervisor states & 570 & 444,816 \\
Supervisor transitions & 1,676 & 3,828,328 \\
Centralized synthesis time (s) & -- & 117.70 \\
\hline
\end{tabular}
\end{table}
\vspace{-1mm}

Table~\ref{tab:centralized_comparison} quantifies the reduction in supervisor size achieved by the hierarchical DES formulation for R1. The hierarchical supervisor reduces the number of states by $\sim 780\times$ and the number of transitions by a $\sim 2300$ compared to the  centralized supervisor. This reduction directly lowers memory requirements and simplifies depth‑first search traversal during online optimization. The centralized synthesis time of 117.70 seconds, even for the smaller R1 scenario, indicates that the monolithic approach becomes infeasible as the map size and the number of UAVs increase. The proposed solution avoids this limitation, bypassing the combinatorial explosion in monolithic DES coordination through a hierarchical approach.

\vspace{-1mm}
\begin{table}[htbp]
\centering
\scriptsize
\caption{Experimental metrics.} \vspace{-2mm}
\label{tab:runtime_metrics}
\begin{tabular}{|l|c|c|}
\hline
\textbf{Metric} & \textbf{R1} & \textbf{R2} \\
\hline
UAVs / nodes / corridors & 2 / 9 / 30 & 4 / 19 / 68 \\
Completed / assigned missions & 4 / 4 & 5 / 6 \\
Mean mission time (s) & 69.11 & 44.42 \\
Maximum mission time (s) & 84.42 & 53.25 \\
Throughput (missions/min) & 1.29 & 2.46 \\
Accepted UTM grants & 32 & 205 \\
Prohibited events: mean / max & 7.03 / 12 & 19.28 / 27 \\
\hline
\end{tabular}
\end{table}
\vspace{-1mm}

The metrics in Table~\ref{tab:runtime_metrics} demonstrate that the hierarchical architecture maintains efficient operation despite the increased complexity. Mean mission time decreases from 69.11 seconds in R1 to 44.42 seconds in R2, while throughput rises from 1.29 to 2.46 missions per minute. Accepted UTM grants increase from 32 to 205, confirming higher service capacity. Although the mean number of prohibited events grows from 7.03 to 19.28, indicating more active constraint enforcement, mission execution times remain bounded and conflict-free coordination is preserved. This behavior stems from the separation between local predictive decision-making and global supervisory restriction: each UAV solves its own MPSC problem, and the UTM layer communicates only the prohibited-event set required to maintain global admissibility.

From the ROS implementation perspective, dedicated nodes were implemented for UTM supervision, UAV execution, mission publication, and task allocation. The UTM node receives discrete events from the fleet, updates the global supervisory state, and broadcasts the prohibited‑event set $\mathcal{P}$. Each UAV node executes a local MPSC loop, interprets admissible events according to the current task and battery condition, and publishes the selected controllable event. Mission management nodes publish delivery requests and coordinate distributed task claiming, avoiding a centralized motion planner. Communication among these nodes was realized through the following ROS topics:

\begin{itemize}
\item \texttt{/event}: System-wide discrete event communication
\item \texttt{/task\_todo}: Administrative task publication  
\item \texttt{/task}: Fleet-wide task allocation
\item \texttt{/prohibited\_events}: UTM global constraint broadcast
\item \texttt{/task\_claims}: Distributed task claiming
\end{itemize}

This node‑topic organization separates continuous vehicle execution, discrete‑event supervision, and mission allocation. Consequently, each UAV solves its local MPSC problem independently, while the UTM layer preserves global admissibility by broadcasting only the prohibitions required to prevent conflicts, enforce geofencing, and maintain mutual exclusion over shared airspace resources.

%% file: sections_V2/07Conclusao.tex
\label{sec:conclusion}

This paper introduced a novel Hierarchical MPSC framework that integrates SCT guarantees with MPC optimization, ensuring collision-free operation and computational tractability for large UAV fleets. The system demonstrated dynamic task allocation and real-time constraint enforcement, confirming suitability for urban air mobility. Future work will extend cost adaptation to the central UTM coordinator, investigate distributed multi-agent coordination, develop map-agnostic decision abstraction, create a federated UTM architecture for metropolitan-scale traffic, and design a UTM task-assignment policy that accounts for UAV characteristics such as battery state and position. Moreover, future work will address the systematic assignment of cost vectors to states and specifications, the treatment of critical low-battery scenarios, and the prioritization of battery constraints over non-critical tasks.

%% file: ifacconf.bib
@book{cassandras2008introduction,
  title={Introduction to discrete event systems},
  author={Cassandras, Christos G and Lafortune, St{\'e}phane},
  year={2008},
  publisher={Springer}
}

@article{ramadge1987supervisory,
  title={Supervisory control of a class of discrete event processes},
  author={Ramadge, Peter J and Wonham, W Murray},
  journal={SIAM journal on control and optimization},
  volume={25},
  number={1},
  pages={206--230},
  year={1987},
  publisher={SIAM}
}

@inproceedings{fahrenberg2011energy,
  title={Energy games in multiweighted automata},
  author={Fahrenberg, Uli and Juhl, Line and Larsen, Kim G and Srba, Ji{\v{r}}{\'\i}},
  booktitle={Theoretical Aspects of Computing -- ICTAC 2011},
  pages={95--115},
  year={2011},
  organization={Springer Berlin Heidelberg},
isbn="978-3-642-23283-1"
}

@article{vilela2022hierarchical,
  title={Hierarchical planning in a supervisory control context with compositional abstraction},
  author={Vilela, Juliana and Hill, Richard},
  journal={Discrete Event Dynamic Systems},
  volume={32},
  number={1},
  pages={89--113},
  year={2022},
  publisher={Springer}
}

@article{lopes2016supervisory,
  title={Supervisory control theory applied to swarm robotics},
  author={Lopes, Yuri K and Trenkwalder, Stefan M and Leal, Andr{\'e} B and Dodd, Tony J and Gro{\ss}, Roderich},
  journal={Swarm Intelligence},
  volume={10},
  number={1},
  pages={65--97},
  year={2016},
  publisher={Springer}
}

@article{roszkowska2013distributed,
  title={A distributed protocol for motion coordination in free-range vehicular systems},
  author={Roszkowska, Elzbieta and Reveliotis, Spyros},
  journal={Automatica},
  volume={49},
  number={6},
  pages={1639--1653},
  year={2013},
  publisher={Elsevier}
}

@techreport{undertaking2017european,
author      = {{SESAR Joint Undertaking}},
  title       = {European Drones Outlook Study: Unlocking the value for Europe},
  institution = {SESAR Joint Undertaking},
  year        = {2016},
  month       = {November},
  address     = {Brussels, Belgium},
  type        = {Study},
  publisher   = {European Union},
  url         = {https://www.sesarju.eu/}
}

@article{xu2020recent,
  title={Recent research progress of unmanned aerial vehicle regulation policies and technologies in urban low altitude},
  author={Xu, Chenchen and Liao, Xiaohan and Tan, Junming and Ye, Huping and Lu, Haiying},
  journal={Ieee Access},
  volume={8},
  pages={74175--74194},
  year={2020},
  publisher={IEEE}
}

@incollection{aweiss2018unmanned,
  title={Unmanned aircraft systems (UAS) traffic management (UTM) national campaign II},
  author={Aweiss, Arwa S and Owens, Brandon D and Rios, Joseph and Homola, Jeffrey R and Mohlenbrink, Christoph P},
  booktitle={2018 AIAA Information Systems-AIAA Infotech@ Aerospace},
  pages={1727},
  year={2018}
}

@article{yadav2021uav,
  title={A uav traffic management system for india: Requirement and preliminary analysis},
  author={Yadav, Anshul and Goel, Salil and Lohani, Bharat and Singh, Shubhanshi},
  journal={Journal of the Indian Society of Remote Sensing},
  volume={49},
  number={3},
  pages={515--525},
  year={2021},
  publisher={Springer}
}

@inproceedings{labib2019multilayer,
  title={A multilayer low-altitude airspace model for UAV traffic management},
  author={Labib, Nader S and Danoy, Gr{\'e}goire and Musial, Jedrzej and Brust, Matthias R and Bouvry, Pascal},
  booktitle={Proceedings of the 9th ACM Symposium on Design and Analysis of Intelligent Vehicular Networks and Applications},
  pages={57--63},
  year={2019}
}

@article{kobayashi2012deterministic,
  title={Deterministic finite automata representation for model predictive control of hybrid systems},
  author={Kobayashi, Koichi and Imura, Jun-ichi},
  journal={Journal of Process Control},
  volume={22},
  number={9},
  pages={1670--1680},
  year={2012},
  publisher={Elsevier}
}

@article{million2007hadamard,
  title={The hadamard product},
  author={Million, Elizabeth},
  journal={Course Notes},
  volume={3},
  number={6},
  pages={1--7},
  year={2007}
}

@book{camacho2007model,
  title={Model Predictive Control},
  year={2007},
  author={Camacho, EF and Bordons, C},
  publisher={Springer}
}

@incollection{bundy1984breadth,
  title={Breadth-first search},
  author={Bundy, Alan and Wallen, Lincoln},
  booktitle={Catalogue of artificial intelligence tools},
  pages={13--13},
  year={1984},
  publisher={Springer}
}

@article{cai2010supervisor,
  title={Supervisor localization: a top-down approach to distributed control of discrete-event systems},
  author={Cai, Kai and Wonham, W Murray},
  journal={IEEE Transactions on Automatic Control},
  volume={55},
  number={3},
  pages={605--618},
  year={2010},
  publisher={IEEE}
}

@article{liu2019scalable,
  title={On scalable supervisory control of multi-agent discrete-event systems},
  author={Liu, Yingying and Cai, Kai and Li, Zhiwu},
  journal={Automatica},
  volume={108},
  pages={108460},
  year={2019},
  publisher={Elsevier}
}

@inproceedings{quigley2009ros,
  title={ROS: an open-source Robot Operating System},
  author={Quigley, Morgan and Gerkey, Brian and Conley, Ken and  Faust, Josh and Foote, Tully and Leibs, Jeremy and Wheeler, Rob and Ng, Andrew Y and others},
  booktitle={ICRA workshop on open source software},
  volume={3},
  number={3.2},
  pages={5},
  year={2009},
}

@article{alves2023ultrades,
  title={Ultrades project-a multiplatform discrete event systems tool},
  author={Alves, Lucas VR and Pena, Patr{\'\i}cia N},
  journal={IFAC-PapersOnLine},
  volume={56},
  number={2},
  pages={6081--6086},
  year={2023},
  publisher={Elsevier}
}

@misc{gurobi2022gurobi,
  title={Gurobi optimizer reference manual, 2023},
  author={Gurobi Optimization, LLC},
  year={2023}
}

@inproceedings{kobetski2006scheduling,
  title={Scheduling of discrete event systems using mixed integer linear programming},
  author={Kobetski, Avenir and Fabian, Martin},
  booktitle={2006 8th International Workshop on Discrete Event Systems},
  pages={76--81},
  year={2006},
  organization={IEEE}
}

@article{harary1962determinant,
  title={The determinant of the adjacency matrix of a graph},
  author={Harary, Frank},
  journal={Siam Review},
  volume={4},
  number={3},
  pages={202--210},
  year={1962},
  publisher={SIAM}
}

@article{loures2025adaptive,
title = {An Adaptive Approach to Multi-Agent Coordination: Leveraging Supervisory Control Theory and MPC for Real-Time Scheduling},
journal = {IFAC-PapersOnLine},
volume = {59},
number = {20},
pages = {2435-2440},
year = {2025},
note = {23th IFAC Symposium on Automatic Control in Aerospace 2025},
issn = {2405-8963},
author = {M.P. Loures and M.A. Santos and L.C.A. Pimenta and P.N. Pena and G.V. Raffo},
}

@article{dulce2022distributed,
  title={Distributed supervisory control for multiple robot autonomous navigation performing single-robot tasks},
  author={Dulce-Galindo, JA and Santos, Marcelo A and Raffo, Guilherme V and Pena, Patricia N},
  journal={Mechatronics},
  volume={86},
  pages={102848},
  year={2022},
  publisher={Elsevier}
}

@techreport{ICA_100_40_Brasil,
  author       = {{Ministério da Defesa - Comando da Aeronáutica}},
  title        = {Aeronaves Não Tripuladas e o Acesso ao Espaço Aéreo Brasileiro},
  institution  = {Departamento de Controle do Espaço Aéreo (DECEA)},
  type         = {ICA 100-40},
  year         = {2020},
  address      = {Brasília, Brazil},
}
